\documentclass[doublecol]{epl2}
\usepackage{amsmath}

\title{Ground state structure and interactions between dimeric
2D Wigner crystals}

\shorttitle{Ground state of dimeric Wigner crystals}
\author{V. Lobaskin \and R. R. Netz}

\institute{
  Physics Department, Technische Universit\"at
  M\"unchen, James-Franck-Str., D-85747 Garching, Germany
 }
\pacs{82.70.-y}{Disperse systems; complex fluids}

\pacs{64.70.Kb}{Solid-solid transitions}

\pacs{68.47.Pe}{Langmuir-Blodgett films on solids; polymers on
surfaces; biological molecules on surfaces}

\abstract{ We study the ground state ordering and interactions
between two two-dimensional Wigner crystals on neutralizing
charged plates by means of computer simulation. We consider
crystals formed by (i) point-like charges and (ii) charged dimers,
which mimic the screening of charged surfaces by elongated
multivalent ions such as aspherical globular proteins, charged
dendrimers or short stiff polyelectrolytes. Both systems, with
point-like and dimeric ions, display five distinct crystalline
phases on increasing the interlayer distance. In addition to
alteration of translational ordering within the bilayer, the phase
transitions in the dimeric system are characterized by alteration
of orientational ordering of the ions. }

\begin{document}
\maketitle

The interaction between electric double layers attracted much
attention in the past twenty years in particular due to the
rediscovered role of ionic correlations. It has been known for
long time that two similarly and strongly charged plates can
attract each other in the presence of multivalent counterions.
This has been seen in Monte Carlo simulations
\cite{Guldbrand,PRL,PRE}, observed experimentally with the surface
force apparatus \cite{73}, deduced from the scattering experiments
on laponite dispersions \cite{Li}, phase diagrams of charged
lamellar systems \cite{39,43}, clay platelets, etc. (see
\cite{Levin,Grosberg1,review} for a review).

As the attraction appears on increasing counterion correlations,
the phenomenon can be conveniently characterized by an
electrostatic coupling parameter $\Xi = 2 q^3 l_B^2 \sigma_s$,
which depends on the Bjerrum length $l_B$, the counterion valency
$q$, and the surface charge density $\sigma_s$. Several successful
descriptions of the attraction have been built for the strong
coupling limit $\Xi \to \infty$, where the correlations are so
strong that the ions form a Wigner crystal (WC) \cite{Netz} or at
least a strongly correlated liquid (SCL) at the charged colloid
surface \cite{50,51}.

Although the phase diagram of a bilayer Wigner crystal has been
known for some time both at the ground state and at finite
temperature \cite{33,33d,structure,33a,33b}, the interaction
between the crystalline double layers as a function of their
separation has not been discussed in detail. The mono- and bilayer
Wigner crystal structures have been addressed in literature in
relation to electrons above the surface of liquid helium,
two-dimensional semiconductor heterostructures, Mott insulators,
dusty plasmas, laser-beam-cooled trapped-ion plasmas, or
dislocation-mediated melting transitions \cite{hetero,swlc1}. In
contrast, in the soft-matter and biological literature the crystal
structure is typically regarded as an auxiliary question for
evaluating the interactions between the surfaces that host these
Wigner crystals \cite{Levine,Lau,Travesset}. Typically, only the
staggered hexagonal crystal, which wins at large distances, is
considered. A few recent publications discuss interaction effects
such as plasmon oscillations using perturbation schemes with
respect to the ground state of the double layers at large
distances (the staggered hexagonal phase) \cite{Levine,Lau}.

A characteristic feature that differentiates Wigner crystals in
soft matter systems from the low-temperature electronic ones stems
from the nature of the ions: The ions have to be multivalent to be
well ordered in aqueous dispersions. The high ion valency also
implies that the Coulomb contribution dominates the free energy
and in-layer thermal fluctuations are negligible. A situation
close to the low-temperature behaviour can be obtained, for
example, when polyelectrolyte molecules, globular proteins or
charged dendrimers play the role of counterions
\cite{Bouyer,Lin,Turesson}. In the general case, such ions are
neither point-like nor spherical and their shape might influence
the interaction between the bilayers. For example, in the limit of
very long ions (DNA chains between lipid bilayers) their
anisotropy leads to orientational ordering and formation of
two-dimensional smectic phases \cite{Salditt,Ohern}. In this work,
we present an accurate numerical solution of the ground state
problem both for point-like (spherical) and small but elongated
(non-spherical) counterions. We first address the interaction
between the bilayers with point-like ions and then use it as a
reference system to study the effect of ion size and charge
distribution on the bilayer interaction.

\begin{figure}
\includegraphics[clip, width=0.48\textwidth]{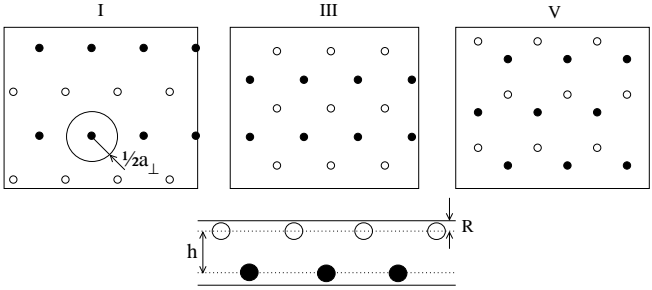}
\caption{Top and side view of three 2D configurations of
counterions on two parallel charged plates. The open and filled
symbols designate the ions located on the opposite surfaces. Set I
corresponds to a single hexagonal lattice, Set III to a
superposition of two square lattices, and Set V corresponds to two
staggered hexagonal lattices (the notation is explained in the
text).} \label{fig:conf}
\end{figure}

We envision a situation where the interaction energy for two flat
parallel Wigner crystals on a neutralizing background is measured
as a function of the interlayer distance. For point-like ions the
scenario includes two obvious limiting cases: at large
separations, the crystals do not feel any discreteness of each
other so that each of them has a 2D plane-filling hexagonal
symmetry. At the smallest separation between them, where both of
them lie in the same plane, a single hexagonal crystal is formed.
At a finer resolution, the transition between these two phases
gives rise to the following sequence of structures on increasing
the interlayer distance: a monolayer hexagonal lattice (I), a
staggered rectangular lattice (II), a staggered square lattice
(III), a staggered rhombic lattice (IV), and a staggered hexagonal
lattice (V) \cite{33}. Here, we evaluate the ground state
interaction by minimizing the energy at each gap width and in
addition, consider three particular phases in more detail. Namely,
we consider three setups, for which the positions of ions on one
plane fall in the geometrical centres of the primitive cells of
the identical lattice on the other plane: the hexagonal monolayer
structure (I); staggered square lattice (III); and staggered
hexagonal lattice (V). These three configurations are shown in
Fig. \ref{fig:conf}. They represent the three rigid lattices on
the bilayer phase diagram, meaning that their structure does not
change within their region of stability. The intermediate
structures (II and IV) are "soft", so that the aspect ratio of
their primary cell is varying with the interlayer distance
\cite{33}. Finally, to study the effect of the ion shape we
replace each ion by a dimer consisting of two identical charges
connected by a stiff spring. The dimer is allowed to rotate and
translate in plane and two different spring lengths are studied.

The energy of a Wigner crystal on a neutralizing charged plane can
be written as
\begin{equation}
 U_1 = \sum_{j=1}^{N-1} \sum_{k=j+1}^{N} \frac{q^2}{ \epsilon
\left| \mathbf{r}_j - \mathbf{r}_k \right|} + \sum_{j=1}^N \frac{2
\pi q_j \sigma_s R}{\epsilon}
\end{equation}
where $\sigma_s$ is the surface charge density of the plate, $q$
the counterion charge, $N$ the number of counterions, $\epsilon$
the dielectric permittivity of the medium, and $R$ the ion-plate
contact distance, or the ion radius for hard sphere ions. The
characteristic lengthscale of the Wigner crystal on a neutralizing
background is the mean lateral distance between the ions
$a_{\bot}$, which is defined by  $\pi \left( a_{\bot}/2 \right)^2
= q /\sigma_s $. The energies can be conveniently expressed in
terms of the average ion-ion Coulomb energy $q^2/( \epsilon
a_{\bot})$: $\tilde{U} = U \epsilon a_{\bot} / q^2 =  4 U \epsilon
/ ( \pi a_{\bot} q \sigma_s) $. In this rescaled form, the result
would explicitly depend on neither ion valency nor the surface
charge density. If we also suppose that the ions form a perfect
crystal so that each ion has exactly the same environment, one
summation over all ions can be performed right away. Thus, we get
for the energy per ion
\begin{equation}
\frac{\tilde{U}_1}{N} = a_{\bot} \sum_{j=2}^{N} \frac{1}
{\left|\mathbf{r}_j - \mathbf{r}_1 \right| } - \frac{8
R}{a_{\bot}}
\end{equation}

For two such plates, in case they are parallel to each other and
have identical ion configurations, the total energy will contain
the self-energy,  2$U_1$, plus the interaction terms: ions --
opposite plate and plate--plate. The ion--plate and plate--plate
energies in the case of charge neutrality will compensate each
other as the sum of distances from each ion of charge $q$ to the
two plates is always $h+2R$, which is exactly equal to the
distance between the corresponding surface element of the same net
charge and the opposite plate. Then the only interesting
contribution comes from the summation over the ions
\begin{equation} \begin{split}
 \frac{\tilde{U}_{12} (h)}{N} &= a_{\bot} \sum_{j=2}^{N}
 \frac{1} {\left|\mathbf{r}_j - \mathbf{r}_1 \right|} \\ &
 +  a_{\bot} \sum_{k=1}^{N}
 \frac{1} {\sqrt{(\mathbf{r}_k -  (\mathbf{r}_1 + \mathbf{a}))^2
 + (h + 2R)^2}
 }  \end{split}
\end{equation}
where $h$ the distance between the ion-containing planes, thus
giving the distance between the plates $h + 2R$, and $\mathbf{a}$
the displacement vector of the lattice on one plate with respect
to the other one. The indices $j$ and $k$ now mean a summation
over the ions on different plates. As the interlayer interaction
leads to structural transformation within the layer, these two
terms remain interconnected and should always be considered
simultaneously. For infinite plates, the total Coulomb energy has
to be calculated numerically. We modeled a piece of 2D crystal of
size $40 \times 40$ to $80 \times 80$ ions with lateral periodic
boundary conditions, and calculated the energies with relative
uncertainty of $10^{-6}$. The plates were supposed to be
homogeneously charged. An MMM2D summation scheme \cite{Axel} and
an MD simulation package ESPResSo v. 1.9 were used
\cite{Espresso}.

\begin{figure}
\begin{center}
\includegraphics[clip, width=0.48 \textwidth]{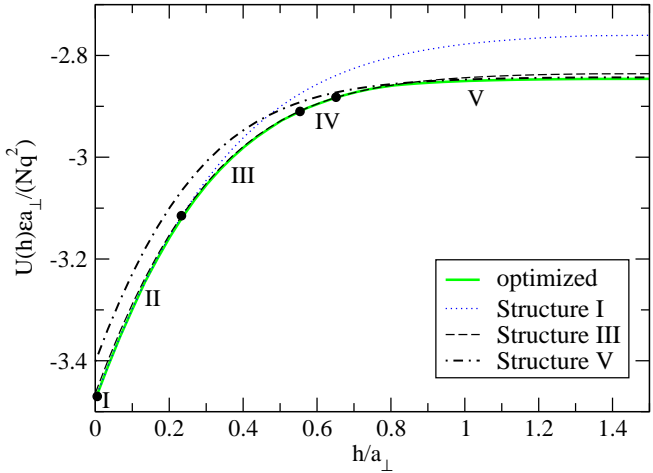}
\end{center}
\caption{Rescaled potential energy per ion of two parallel Wigner
crystals on oppositely charged plates as a function of normalized
distance $h/a_{\bot}$ between the crystals for the three different
counterion arrangements. The optimized structure is obtained by
the energy minimization using a Brownian dynamics simulation at
zero temperature. The transition points as reported in Ref.
\cite{33} are marked by filled circles. The regions of stability
of various crystalline structures are delimited by the circles and
marked next to the optimized energy curve.} \label{fig:e1}
\end{figure}

\begin{figure}
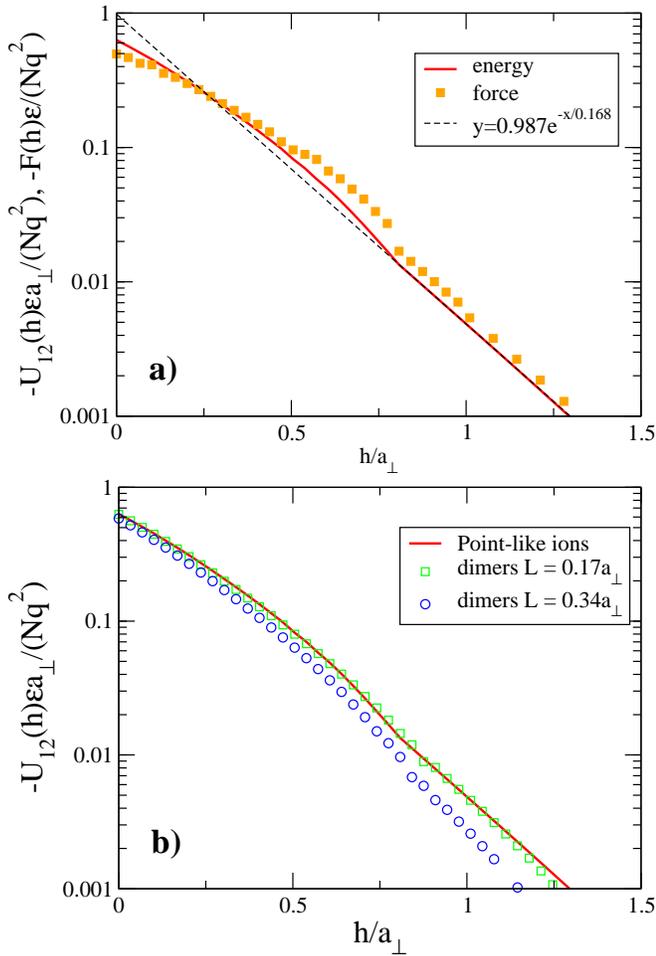

\begin{center}
\includegraphics[clip, width=0.48 \textwidth]{fig3a.eps}
\includegraphics[clip, width=0.48 \textwidth]{fig3b.eps}
\end{center}
\caption{a) Absolute value of rescaled attraction energy and force
between two charged plates with crystalline arrangement of
point-like ions. The dashed line shows the predicted asymptotic
behaviour at large $h/a_{\bot}$ \cite{33}. b) The rescaled
interaction energy between two parallel Wigner crystals formed by
charged dimers of the indicated length in comparison with that for
point-like ions. The dimer charge as well as the number density
are matching those for the system with point-like ions. }
\label{fig:dimer}
\end{figure}

We first consider the relative energies of the preformed lattices.
Figure \ref{fig:e1} shows the potential energy of interaction
between the two layers as a function of the gap width $h$ for the
three setups. The total energy, $U=2U_1 + U_{12}$, in all of them
is strongly negative and decreases in absolute value with the gap
width. The limiting value at $h \to \infty$ on each curve
corresponds to twice the energy of a single Wigner crystal $U_1$.
To calculate the true ground state energy we perform molecular
dynamics simulation of two ionic layers with the ions constrained
to their corresponding planes but allowed to move within the
plane. For each interlayer distance we start simulation from each
of the three preformed lattices and then record the minimal
energy. We see that the minimal energy from the free configuration
coincides with energy of the favorable arrangement in the
appropriate range of distances and gets lower in two intermediate
regions close to the transition points I $\to$ III and III $\to$ V
(Fig. \ref{fig:e1}). In these regions, one should expect an
appearance of the staggered rectangular lattice and the staggered
rhombic one, respectively. At $h=0$ we obtain the rescaled energy
per ion $\tilde{U}_{12}= -3.47502$, which is close to value
$-3.47493$ reported in Ref. \cite{33} (the scaling factor in our
work differs from Ref. \cite{33} by $\sqrt{\pi}$). The single
hexagonal lattice is stable in a very small region close to
$h/a_{\bot} = 0$. The location of the further transition points
can be estimated from comparing the energies of the corresponding
phases. The intersection point of the curves I and III in Fig.
\ref{fig:e1} corresponds to a transition into staggered square
phase. The energy of the optimized structure becomes lower than
that of the phase III at $h/a_{\bot} \approx 0.55$, which
indicates a transition into phase IV. The structure V prevails at
$h/a_{\bot} \geq 0.77$. We note that the transition in our work is
observed at a higher $h/a_{\bot}$ than it was reported in
\cite{33,structure}. The interaction energy of the two layers is
shown in Fig. \ref{fig:dimer}a. The energy curve is smooth between
the transition points. The force curve shows a jump at $h/a_{\bot}
\approx 0.77$, which is expected due to discontinuous character of
the transition between phases IV and V \cite{33}. The initial
decay of the interaction energy is close to linear, while the
asymptotic behaviour at $h/a_{\bot}> 1$ is clearly exponential.
The observed behaviour is close to the asymptotic decay expected
at large distances $U_{12}(h) \propto \exp (- h/(a_{\bot}/2 \pi))$
\cite{33,50} (see Fig. \ref{fig:e1}).

\begin{figure}
\begin{center}
\includegraphics[clip, width=0.48\textwidth]{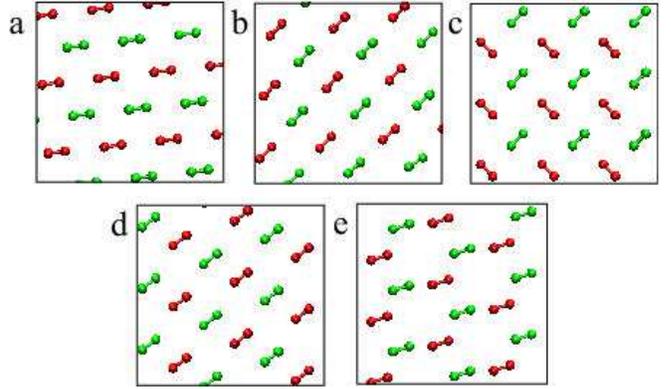}
\end{center}
\caption{Snapshots from simulations of dimeric bilayers with dimer
length $L = 0.34 a_{\bot}$ at different gap widths: $h / a_{\bot}
= 0.1 \texttt{(a)}, 0.33 \texttt{(b)}, 0.5 \texttt{(c)},
0.75\texttt{(d)},1.2 \texttt{(e)}$. The different colors
correspond to ions located at different planes. }
\label{fig:order}
\end{figure}

We now look at the interaction between crystals formed by
Coulombic dimers for two different lateral ion sizes $L = 0.5R =
0.17a_{\bot}$ and $L = R = 0.34 a_{\bot}$. As an additional degree
of freedom is involved, we expect a richer phase behaviour in this
case. In the limit $L=0$ the behaviour of the system with
point-like ions is recovered. Figure \ref{fig:dimer}b presents the
interaction energies as a function of the gap width. Simulation
snapshots for three different values of $h/a_{\bot} = 0.1, 0.33,
0.5, 0.75$ and $1.2$, which correspond to different crystalline
structures, are presented in Figure \ref{fig:order}. We see that
at small and large relative distances the dimers in the two layers
tend to be aligned while at the intermediate separations the
dimers in each layer tend to orient perpendicular to their nearest
neighbors in the 2D projection of the lattice (either same plane
neighbors or the neighbors in the opposite plane).

A variation of the dimer length affects only the ion-ion part of
the total energy of the bilayer. A system with longer dimers has a
higher self-energy of each layer while the corresponding
interlayer part depresses the interaction. We see from Fig.
\ref{fig:dimer}b that the interaction between the layers indeed
becomes weaker on increasing the dimer length. Roughly, the
characteristic length $a_{\bot}$ decreases by $L/2$ as compared to
the point charges. Then, the interaction energy would decay as
$U(h) \propto \exp (2 \pi h/(a_{\bot}+ L/2))$. A fit to the
calculated interaction energies reveals the decay length change
from $0.19 a_{\bot} $ for the point-like ions to $0.15 a_{\bot}$
for the system with $L = 1$ (or $0.34 a_{\bot}$). Ultimately, at
$L = a_{\bot}$ at small separations one should recover the bilayer
with $ a'_{\bot} = a_{\bot}/\sqrt{2}$.

In simulations with short dimers with $L=0.17 a_{\bot}$ at various
gap thicknesses we observe: (i) $0 < h/a_{\bot} < 0.27$ a
staggered parallelogrammetic lattice, parallel orientation of
neighboring dimers, soft; (ii) $0.27 < h/a_{\bot} < 0.8$ staggered
square lattice, perpendicular dimer orientation, rigid; (iii) $0.8
< h/a_{\bot}$ staggered rhombic lattice, parallel dimer
orientation, rigid. At $L=0.34 a_{\bot}$, we have the following
sequence of structures: (i) $0 < h/a_{\bot} < 0.17$ a staggered
rectangular lattice, parallel orientation of neighboring dimers,
rigid, the longer lattice constant $a_2$ is roughly $a_1 + L$
(Fig. \ref{fig:order}a); (ii) $0.17 < h/a_{\bot} < 0.40$ we
observe domains of rhombic lattice, rotated with respect to the
neighbouring domains (Fig. \ref{fig:order}b); (iii) $0.40 <
h/a_{\bot} < 0.67$ staggered square lattice, perpendicular dimer
orientation, rigid (Fig. \ref{fig:order}c); (iv) $0.67 <
h/a_{\bot} < 0.91$ staggered parallelogrammetic lattice, parallel
dimer orientation, rigid (Fig. \ref{fig:order}d); (v) $0.91 <
h/a_{\bot}$ staggered triangular (a honeycomb-like) lattice,
parallel, rigid (Fig. \ref{fig:order}e). We note that in contrast
with some phases of point ions (the "soft" rectangular (II) and
rhombic (IV) phases), the observed lattices in the system with
long dimers ($L=0.34 a_{\bot}$) are rigid, i.e. the aspect ratio
of the primitive cell stays constant within the region of
stability of the corresponding lattice.

\begin{figure}
\begin{center}
\includegraphics[clip, width=0.48\textwidth]{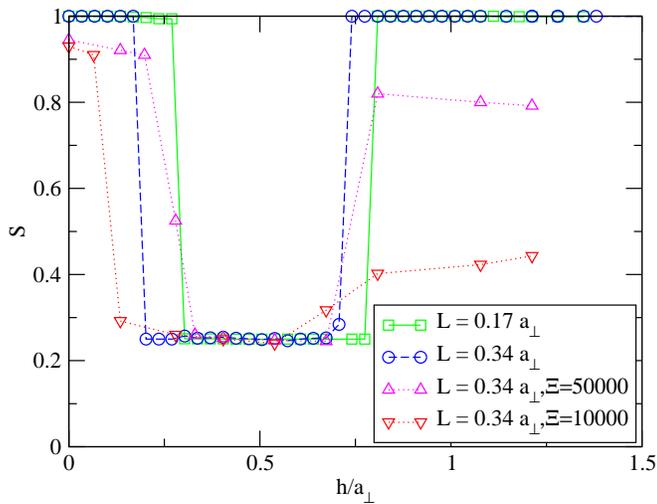}
\end{center}
\caption{ Orientational order parameter for dimeric Wigner
bilayers with two different dimer sizes. The value $S=1$
corresponds to perfectly aligned dimers, $S = 0.25$ to a
coexistence of two perpendicular dimer orientations. The lines are
guide to the eye.} \label{fig:order1}
\end{figure}

The most interesting observation for the dimeric systems is
related to the ability of dimers to adjust the orientation to
minimize the electrostatic energy. The effect is strongest for the
square lattices where we find perpendicular orientation of the
neighboring dimers. A similar phenomenon of altering orientational
ordering was discussed recently for a one-dimensional stack of
rod-like ions \cite{Fazli}. In contrast to our observation for 2D
layers, the ground state in a staggered 1D system is represented
by perpendicular orientation of the neighboring ions, which then
changes to a twisted chiral phase on increasing density. In our
system, the reorientation of the ions happens suddenly on varying
the separation distance between the surfaces, which might find an
application in nano-structuring. This structural transformation
can be best characterized by the scalar order parameter $S =
\frac{1}{2}\langle 3 \cos^2 \theta - 1\rangle$ (the second moment
of the  orientational distribution function), where $\theta$ is
the angle between the molecule orientation and the director, and
average over all dimers is taken. This value is plotted in Fig.
\ref{fig:order1}. One can see that the phases at short and long
interlayer distances have $S=1$, which corresponds to ideally
aligned dimers (Fig. \ref{fig:order}a,d,e). Further on, the $S$
values of $0.25$ correspond to a coexistence of two preferred
dimer orientations (Fig. \ref{fig:order}c), which are
perpendicular to each other. The arrangements with two preferred
orientations are observed in the region of stability of the square
lattice. From the plot shown in Fig. \ref{fig:order1}, we see that
the onset of the perpendicular dimer orientation as well as the
return into the aligned state happens in the system with longer
dimers at smaller distances.

Finally, we note that the presented sequence of orientational
transitions exists also at finite temperatures. A simulation
performed for $L/a_{\bot}=0.34$ at $\Xi = 50000$ and $\Xi=10000$
still showed three regions with high, then low, and again high
order parameter on increasing the interlayer distance, although
the short-distance and the long-distance phases become less
aligned due to thermal fluctuations. The observed order parameter
was $S \approx 0.9$ at small interlayer separations, and $S
\approx 0.8$ ($\Xi = 50000$) and $S \approx 0.4$ ($\Xi = 10000$)
at large separations (Fig. \ref{fig:order1}, triangles), while the
values for the perpendicular dimer orientations, $S \approx 0.25$,
did not change with temperature. One can expect that the
temperature region of stability of the square lattice with the
perpendicular dimer orientation is larger than that for the
aligned phases, which follows from the increasing stability of the
square lattice itself in the system with point ions \cite{33a}. We
also note that the bilayer system is stable with respect to normal
fluctuations, which are suppressed in our setup by the repulsion
between the mobile ions that confines the ions to thin layers at
the walls.

We calculated the ground state interaction energy of two planar
ionic double layers represented by Wigner crystals of point-like
charges or charged dimers, where the latter model mimics the
situation of screening the interaction between charged surfaces
with polyelectrolytes or non-spherical molecules. The crystalline
structure observed with point-like charges agrees with that
reported in the earlier literature, while novel structures with
unusual orientational ordering were observed in the system with
elongated ions. In all cases we found exponential asymptotic decay
of the correlation attraction at interlayer distances larger than
the characteristic lateral separation between the counterions. We
found that the elongated ions produce in general weaker
correlation attraction attraction than the point-like or spherical
ions of the same charge.

\acknowledgments We are grateful to F. M. Peeters for introducing
us to his works on bilayer Wigner crystals and for his comments on
the manuscript. The work was partly supported by a program
``Material Science of Complex Interfaces'' of Elitenetzwerk
Bayern.

\end{document}